\documentclass[usenatbib,usegraphicx]{mn2e}

\usepackage{subfigure}
\usepackage{longtable}
\usepackage{graphicx}
\usepackage{graphics}
\usepackage{keyval}
\usepackage{trig}
\usepackage[american]{babel}
\usepackage{url}
\usepackage{color}
\usepackage{amsmath}
\usepackage{bm}

\def\beq{\begin{equation}}
\def\eeq{\end{equation}}
\def\bey{\begin{eqnarray}}
\def\eey{\end{eqnarray}}

\def\msun{M_\odot}

\def\lsim{\mathrel{\raise.3ex\hbox{$<$\kern-.75em\lower1ex\hbox{$\sim$}}}}
\def\gsim{\mathrel{\raise.3ex\hbox{$  $\kern-.75em\lower1ex\hbox{$\sim$}}}}

\def\tz{t_\mathrm{0}}
\def\te{t_\mathrm{E}}

\def\uz{u_\mathrm{0}}

\def\thetae{\theta_\mathrm{E}}

\def\ds{D_\mathrm{S}}
\def\dl{D_\mathrm{L}}
\def\pils{\pi_\mathrm{LS}}

\def\a0{A_{\mathrm{0}}}

\def\deltapar{\delta_{\parallel}}
\def\deltaperp{\delta_{\perp}}
\def\tobs{T_{\mathrm{obs}}}

\def\tast{t_{\mathrm{ast}}}

\def\deltat{\delta_{\mathrm{T}}}
\def\mbh{M_{\mathrm{BH}}}
\def\c2dof{\chi^2/{\rm d.o.f.}}
\def\pdet{P_{\rm det}}
\def\muls{\mu_{\mathrm{LS}}}
\def\m606{m_{\rm F606W}}
\def\m814{m_{\rm F814W}}
\def\avn{\langle N_{\rm det} \rangle}
\def\nd{N_{\rm D}}
\def\np{N_{\rm P}}
\def\biclin{\mathrm{BIC_{lin}}}
\def\biclens{\mathrm{BIC_{lens}}}
\def\bicbin{\mathrm{BIC_{bin}}}
\def\dbic{\Delta_\mathrm{BIC}}
\def\dbict{\Delta_{\mathrm{BIC}, T}}
\def\dobs{\delta_{\mathrm{obs}}}

\newcommand\Eq[1]{Eq.~(\ref{#1})}
\newcommand\Fig[1]{Fig.~\ref{#1}}
\newcommand\Tab[1]{Table~\ref{#1}}
\newcommand\Sec[1]{Sec.~\ref{#1}}

\title{Searching for intermediate-mass black holes in globular clusters with gravitational microlensing}

\author[N.~Kains et al.]
{\parbox{\textwidth}{N.~Kains$^{1}$, D.~M.~Bramich$^{2}$, K.~C.~Sahu$^{1}$, A.~Calamida$^3$}\vspace{0.4cm}\\
\parbox{\textwidth}{$^1$Space Telescope Science Institute, 3700 San Martin Drive, Baltimore, MD 21218, United States of America \label{stsci}\thanks{nkains@stsci.edu}\\
$^2$Qatar Environment and Energy Research Institute (QEERI), HBKU, Qatar Foundation, Doha, Qatar \label{qeeri}\\
$^3$National Optical Astronomy Observatory, 950 N Cherry Ave, Tucson, AZ 85719, United States of America \label{noao}\\}}

\begin{document}

   \date{Received ... ; accepted ...}

 
\pagerange{\pageref{firstpage}--\pageref{lastpage}} \pubyear{2015}

\maketitle

\label{firstpage}

  \begin{abstract}

We discuss the potential of the gravitational microlensing method as a unique tool to detect unambiguous signals caused by intermediate-mass black holes in globular clusters. We select clusters near the line of sight to the Galactic Bulge and the Small Magellanic Cloud, estimate the density of background stars for each of them, and carry out simulations in order to estimate the probabilities of detecting the astrometric signatures caused by black hole lensing. We find that for several clusters, the probability of detecting such an event is significant with available archival data from the \textit{Hubble Space Telescope}. Specifically, we find that M~22 is the cluster with the best chances of yielding an IMBH detection via astrometric microlensing. If M~22 hosts an IMBH of mass $10^5\msun$, then the probability that at least one star will yield a detectable signal over an observational baseline of 20 years is $\sim 86\%$, while the probability of a null result is around 14\%. For an IMBH of mass $10^6\msun$, the detection probability rises to $>99\%$. Future observing facilities will also extend the available time baseline, improving the chance of detections for the clusters we consider.

\end{abstract}               

   \begin{keywords}
   globular clusters -- black holes -- intermediate-mass black holes -- gravitational lensing -- microlensing
    \end{keywords}


\section{Introduction}\label{sec:intro}

After formation, a stellar-mass black hole may grow via accretion of surrounding material, or by merging with other black holes; eventually, supermassive black holes (SMBHs) may form with masses ranging upwards of $\sim 10^6\, \msun$. The detection of SMBHs at large redshifts indicates that some of them formed quickly and were already present only a few hundred million years after the Big Bang \citep{fan06}. Explaining how these objects formed so rapidly is a challenge, because a stellar-mass seed black hole cannot reach a mass of $\sim 10^6\, \msun$ within 1 Gyr even by accreting material at the highest possible rate, the Eddington rate, although mechanisms have been put forward that could enable accretion at super-Eddington rates \citep{alexander14}. In this context, one of the preferred scenarios for such rapid initial growth is through the merger of smaller seed black holes of intermediate mass ($10^2$-$10^6\,\msun$, e.g. \citealt{ebisuzaki01}), which serve as the missing link to understanding the growth of SMBHs.

Globular clusters provide dense enough stellar environments for intermediate-mass black holes (IMBHs) to form through runaway mergers of stars (e.g. \citealt{portegies02}, \citealt{miller02}), and they are approximately the same age as their host galaxy, suggesting that the IMBHs required for the growth of SMBHs in the early Universe might have been delivered to galaxy centres by globular clusters \citep[e.g.][]{dolcetta01, lutzgendorf12}. Further motivation for searching for IMBHs in clusters comes from the well-known $M-\sigma$ relation (e.g. \citealt{ferrarese00}) for galaxies, which hints at a fundamental connection between the formation and evolution of central black holes and the central kinematics of galaxies. Extrapolating this relation to lower masses implies that IMBHs should be found in systems with central dispersions of $\sim$10-20 km/s, which are the dispersions typically found in globular clusters. 

Since first proposed by \cite{silk75}, the existence of IMBHs in clusters has been probed with various techniques. Attempts to detect accretion signatures through X-ray and radio observations have generally only yielded upper mass limits that depend on the assumptions made about the accretion process and the density of the surrounding material (e.g. \citealt{grindlay01}; \citealt{maccarone05}; \citealt{haggard13}). In spite of this, some promising IMBH candidates \citep[e.g.][]{farrell12, soria13, mezcua15} have been identified through observations of ultra-luminous X-ray sources (ULXs). These are extra-nuclear point sources with X-ray luminosities greater than $10^{39}$ ergs s$^{-1}$, corresponding to the Eddington limit of a 10$\msun$ black hole \citep{roberts07}.

Surface brightness profiles of globular clusters hosting central IMBHs are expected, from both theoretical predictions and $N$-body simulations, to have weak central cusps \citep[e.g][]{bahcall76, baumgardt05} as opposed to core-collapsed clusters with steep profiles and pre-core collapsed systems with no cusp. However, it has also been shown from numerical simulations that a photometric profile with a shallow cusp might also be a sign of ongoing core collapse \citep{trenti10}. Therefore a weak central cusp is not a unique signature of a central IMBH, making claims of IMBH detections using this method contentious (e.g. \citealt{lanzoni07b, vesperini10}). Furthermore, the presence of anisotropic orbits could mimic the signature of a central IMBH in kinematic profiles, making the interpretation of cusp data ambiguous \citep{ibata09}. Mass segregation of stellar remnants can also replicate such signatures, as shown by \cite{illingworth77} and then by \cite{baumgardt03} in their analysis of the kinematic profile of M~15. Because remnants are natural products of stellar evolution, it is then difficult to favour an IMBH scenario.

Combining photometric and spectroscopic observations to yield kinematic data that can be compared to dynamical models has recently led to a number of claims of IMBH detections, thanks to improvements in instrumental resolution and the use of integral field units (\citealt{lutzgendorf13a}). This method has also yielded a mass estimate for an IMBH in $\omega$ Cen of $(4.7 \pm 1.0) \times 10^4\, \msun$ \citep{noyola10}. However, other authors (e.g. \citealt{anderson10}) have found less compelling evidence for a central black hole in $\omega$ Cen, which is in any case suspected to be the stripped nucleus of a dwarf galaxy rather than a true globular cluster (e.g. \citealt{noyola08}). Different measurements using this technique have also led to conflicting predictions as to the presence of an IMBH in a few clusters \citep[e.g.][]{kamann14}.

Further recent detection claims include the work of \cite{pasham14}, who reported quasi-periodic oscillations in the X-ray emission of ULX M~82 - X-1, which they then used to estimate a black hole mass of $\sim400\msun$. However, the reliability of this type of oscillations to constrain black hole masses has been questioned by other authors \citep[e.g.][]{middleton11}. \cite{baldassare15} estimated a mass of $5\times10^4\msun$ for the black hole in the centre of the dwarf galaxy RGG~118, using virial black hole mass estimate techniques, the limitations and caveats of which are discussed in detail by \cite{shen13}. \cite{oka16} concluded that the velocity dispersion in the molecular cloud CO-0.40-0.22 is best modelled by the gravitational effect of a $10^5\msun$ black hole. The proximity of this molecular cloud to the Milky Way's central SMBH, Sgr A$^*$, is particularly interesting within the context of IMBHs being potential seeds for SMBH formation.

Despite this wealth of indirect observational evidence, there has not yet been an unambiguous detection of an IMBH. In this paper, we discuss how gravitational microlensing would allow us to detect an astrometric signal that could be unambiguously attributed to the presence of an IMBH. While \cite{safonova10} have already proposed using microlensing as a technique to detect IMBHs in cluster cores, they only considered the detection of photometric signals of microlensing of cluster stars by the IMBH, which have extremely low detection probabilities. Here, we will show that astrometric microlensing is a far more promising method to achieve a detection. We conduct a brief review of astrometric and photometric microlensing in \Sec{sec:microlensing}, and how it can be used to measure the mass of single objects (Sec. \ref{sec:massmeas}). The feasibility of such a detection is discussed in \Sec{sec:feasibility}, and we describe simulations to estimate expected event rates and the probabilities of detecting at least one event in several chosen globular clusters in \Sec{sec:simulations}. We discuss our findings in \Sec{sec:discussion}, and draw conclusions for potential future detections in \Sec{sec:conclusions}.

\section{Astrometric and photometric microlensing}\label{sec:microlensing}

Astrometric microlensing has been discussed in detail by \cite{dominik00}. The interested reader is referred to that publication for a full discussion. Here we recall only the essential details.

A microlensing event occurs when the observer, a source at distance $\ds$, and a lens of mass $M$ at a smaller distance $\dl$ become aligned. The time-dependent angular separation $\phi$ of the lens and source is usually expressed in units of the Einstein ring radius as $u=\phi/\thetae$, where

\begin{equation}\label{eq:thetae}
\thetae = \sqrt{\frac{4GM}{c^2}(\dl^{-1} - \ds^{-1})}\, .
\end{equation}

\begin{figure*}
  \centering
  \includegraphics[width=8cm, angle=0]{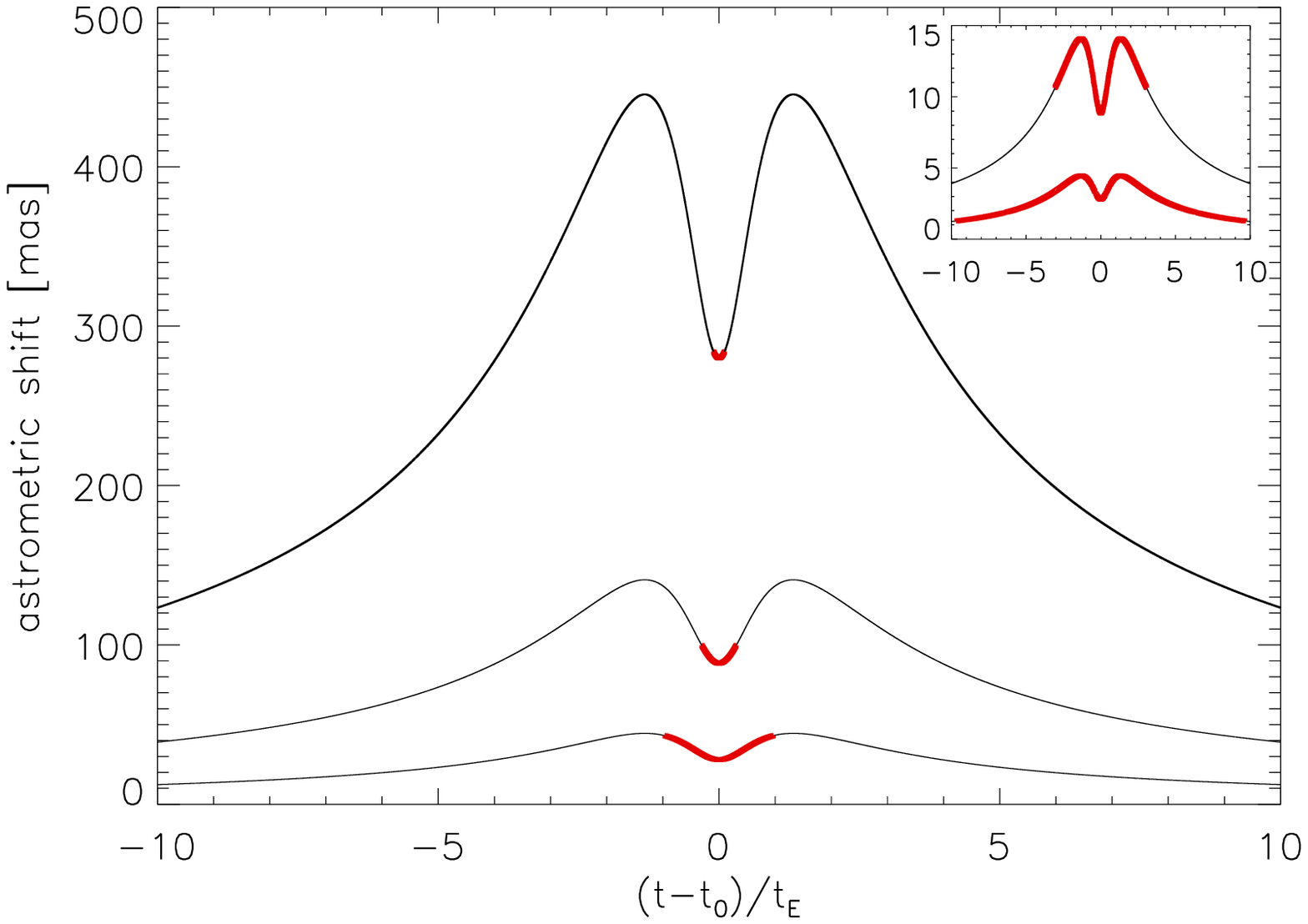}
  \includegraphics[width=8cm, angle=0]{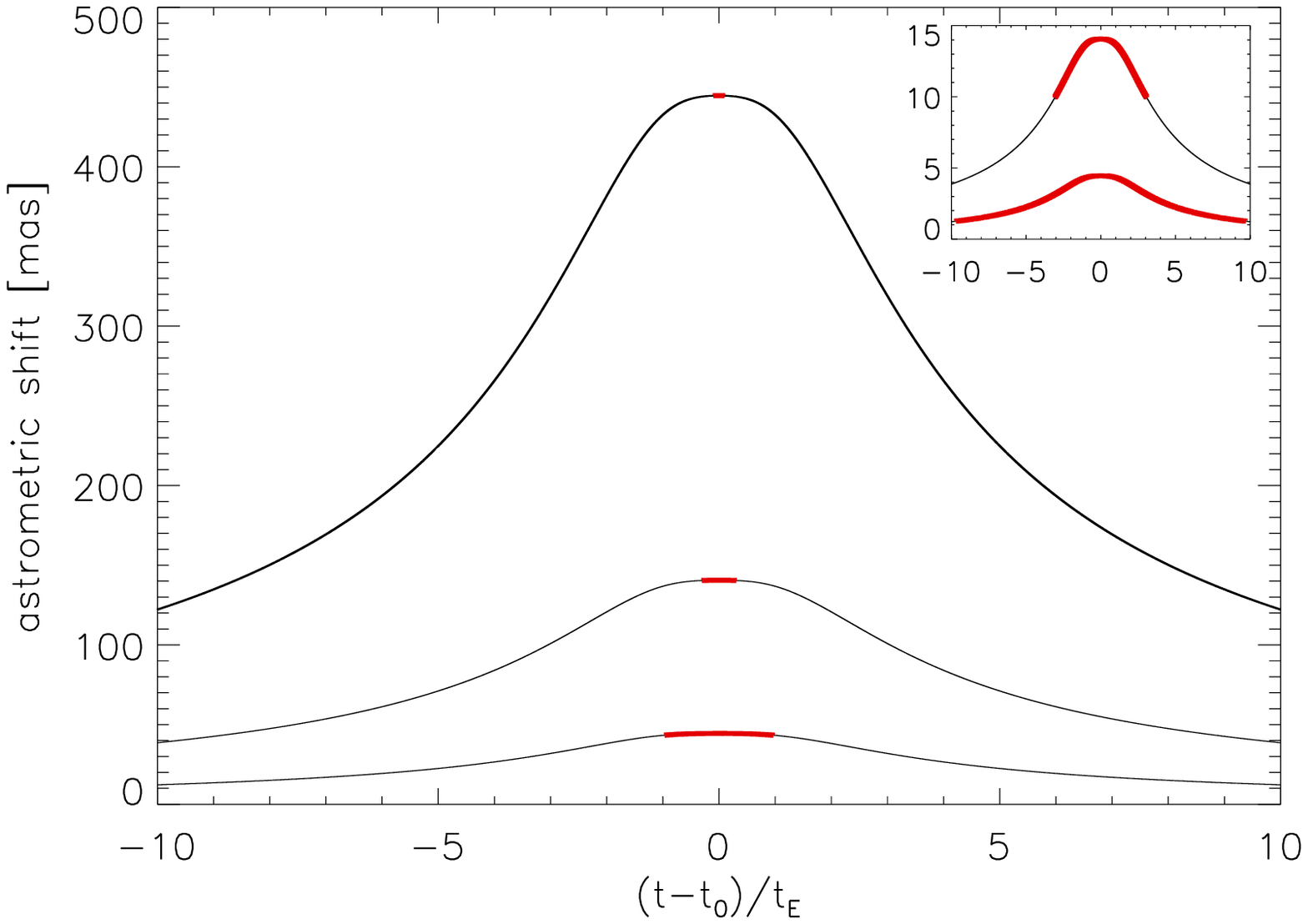}
  \includegraphics[width=8cm, angle=0]{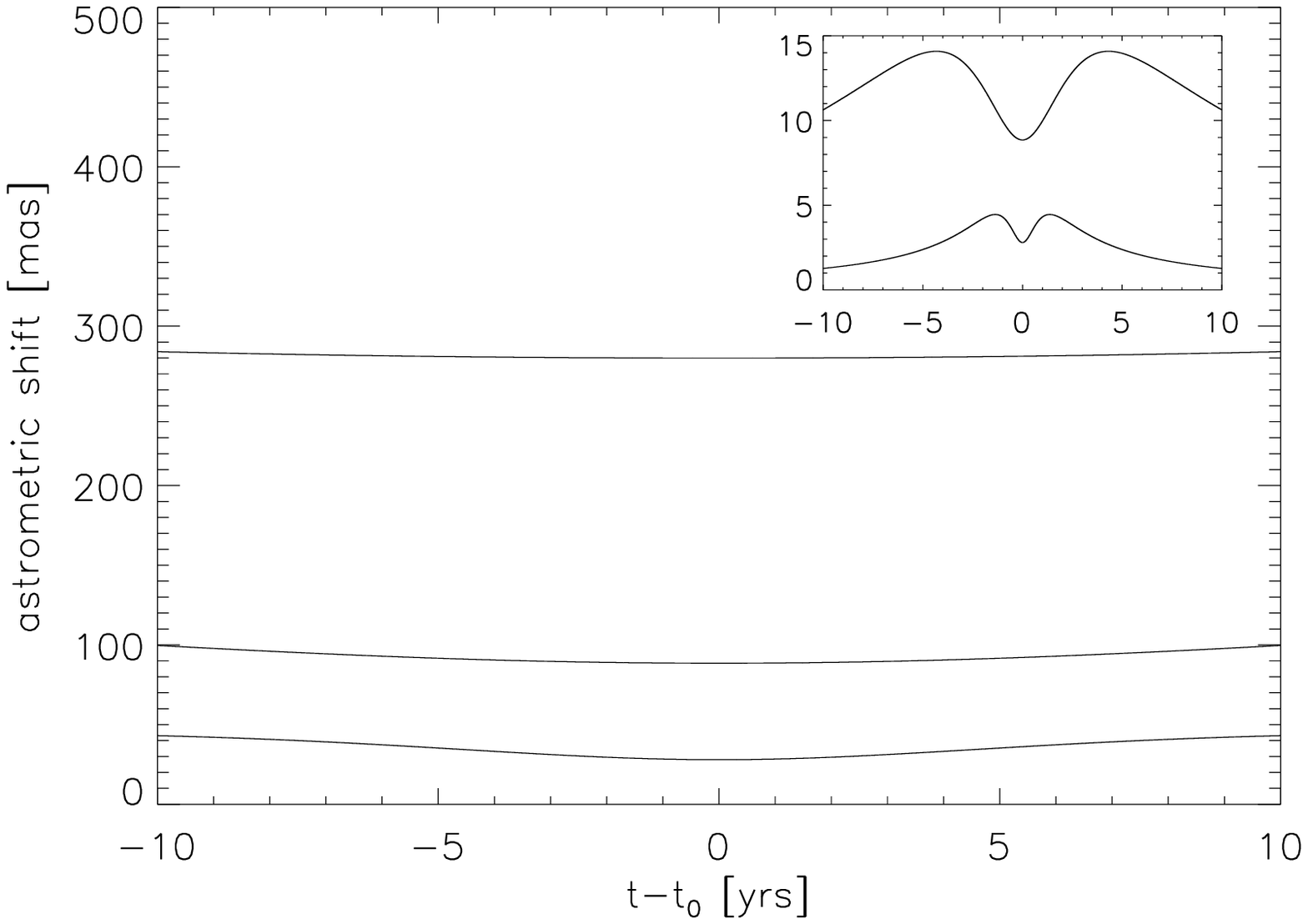}
  \includegraphics[width=8cm, angle=0]{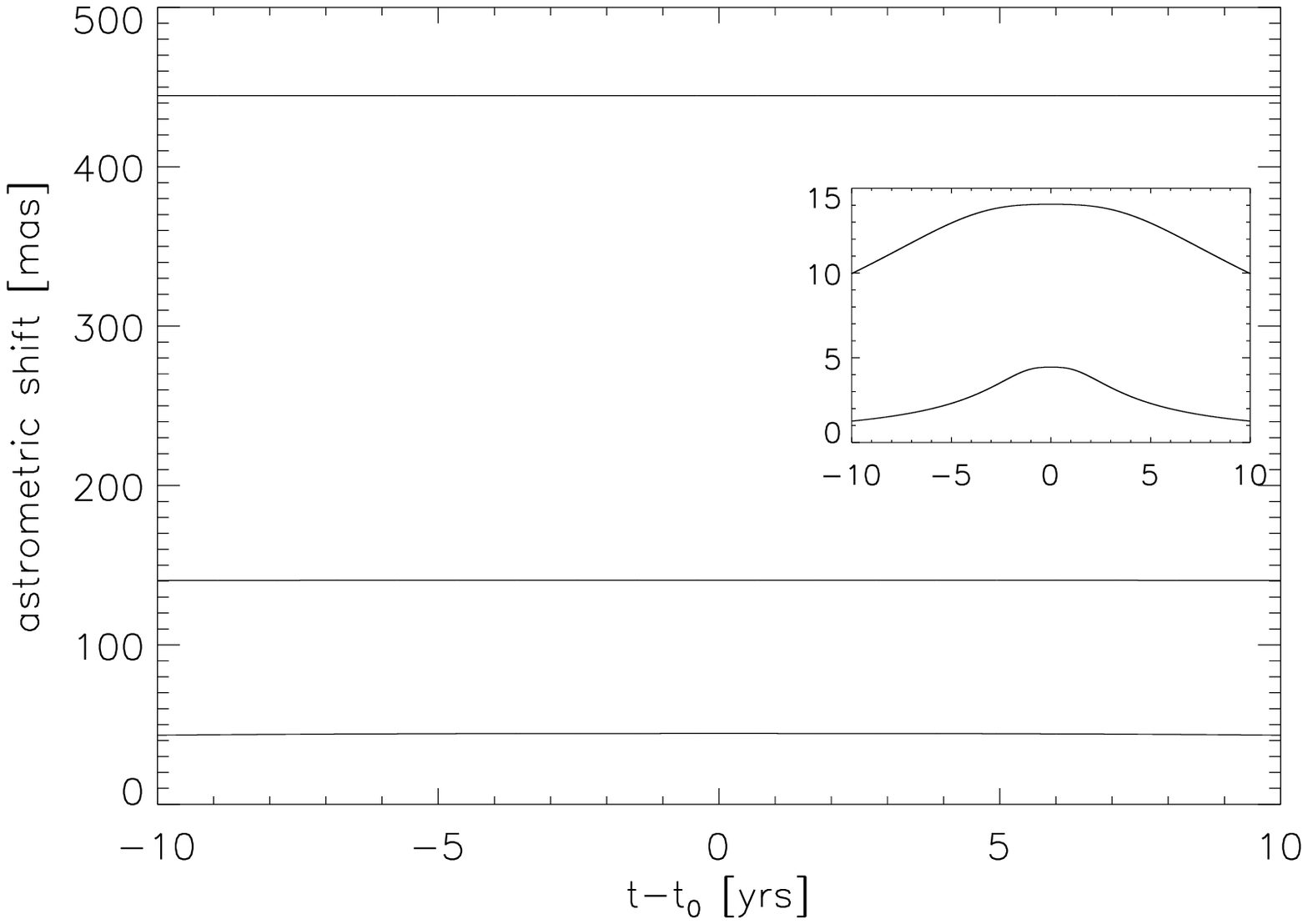}  
  \caption{The time-dependent centroid shift due to lensing by an IMBH of masses $10^6, 10^5, 10^4, 10^3$, and $10^2\,\msun$ plotted (with the signal amplitude increasing with mass, and the lower two masses in the inset) for $\dl=3.2$ kpc, $\ds=8.5$ kpc, $\muls=12.2$ mas/yr (corresponding to a Bulge source and a lens in M~22), $\uz=0.5$ (\textit{left}) and 1.5 (\textit{right}). The upper panels have a time axis in $\te$, while the lower panels show the signal for the various masses over a range of 20 years, with a time axis in years. The red segments in the upper panels indicate the part of the astrometric curve that would be covered by a 20 year campaign centred on $t=\tz$, i.e. the same part that is shown in the corresponding lower panels. \protect\label{fig:imbh_shift_time}}
\end{figure*}

\begin{figure}
  \centering
  \includegraphics[width=8cm, angle=0]{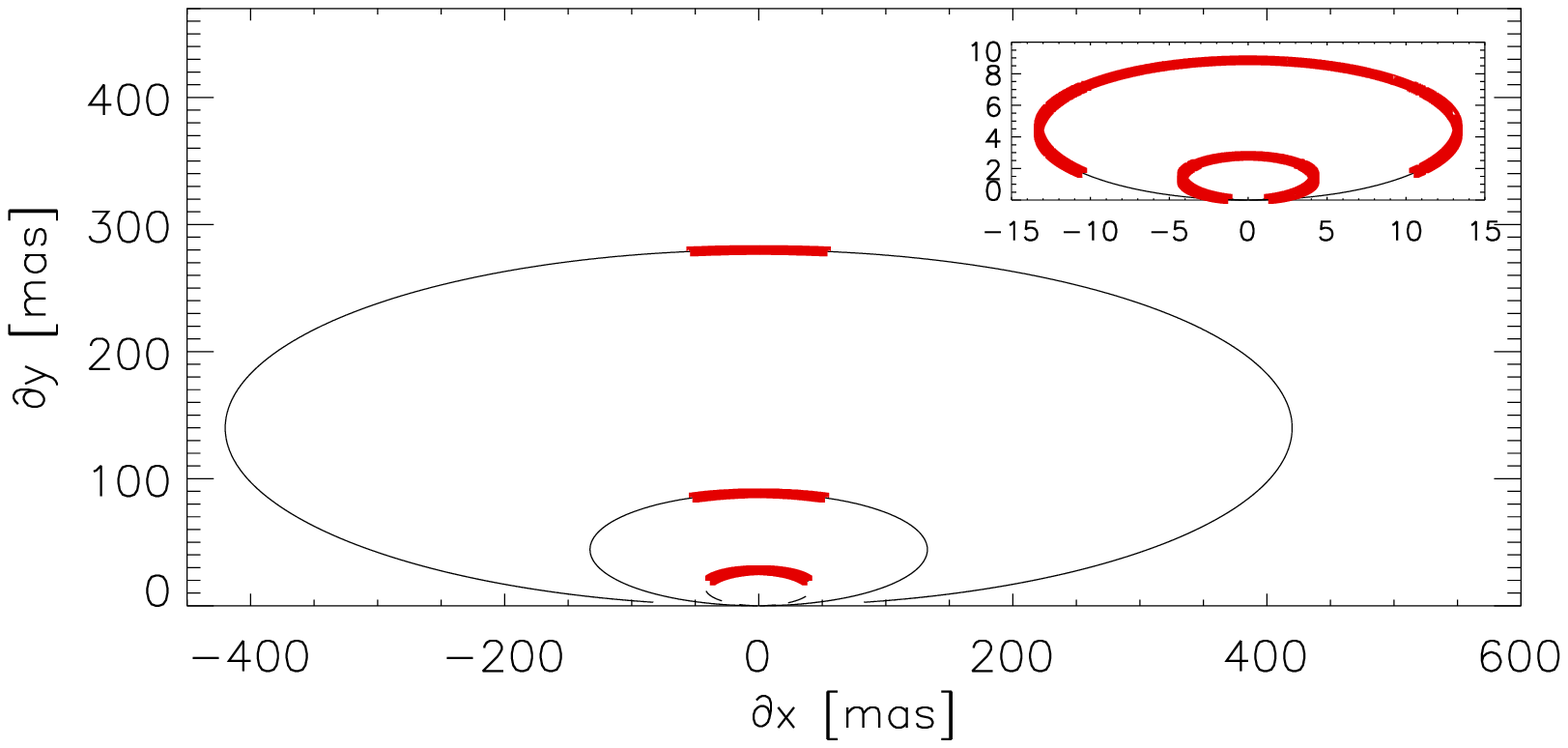}
  \includegraphics[width=8cm, angle=0]{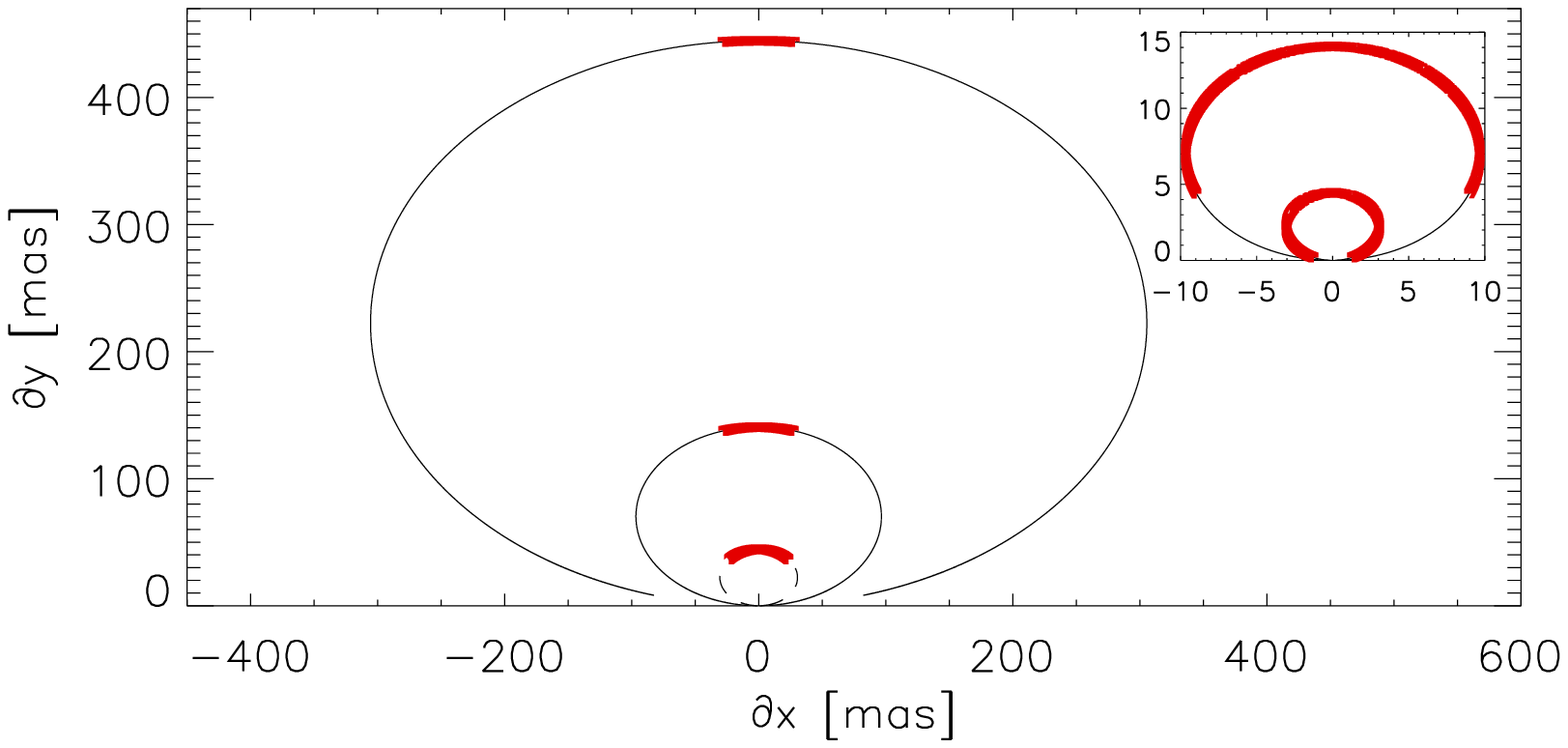}
  \caption{2-D astrometric shift during a lensing event for five lens masses ($10^6, 10^5, 10^4, 10^3$, and $10^2\,\msun$, with the ellipse size increasing with mass, and the two lower masses shown in the inset), with the same IMBH parameters as \Fig{fig:imbh_shift_time}, for $\uz=0.5$ (top) and $\uz=1.5$ (bottom). Overplotted in thick red is the part covered by a 20-year observing campaign centered on $t=\tz$. \protect\label{fig:imbh_shift_pos}}
\end{figure}

The photometric microlensing event then consists in the apparent magnification of the source, due to gravitational deflection of its light rays by the lens. A point source is magnified by a factor \citep[e.g.][]{paczynski86}

\begin{equation}\label{eq:muu}
\mu(u) = \frac{u^2 + 2}{u\sqrt{u^2+4}}\, ,
\end{equation}

\noindent
where $u$, and therefore $\mu(u)$, is time-dependent.
Due to the asymmetric nature of the images of the source produced by the gravitational lens, the apparent position of the source also appears to change with time as the event unfolds following a characteristic pattern; this constitutes the astrometric microlensing. The apparent displacement of the centroid of a point source by an amount $\delta(u)$ can be expressed as \citep{hog95}

\begin{equation}\label{eq:deltau}
\delta(u)=\frac{u}{u^2+2}\thetae\, ,
\end{equation}

\noindent
with the displacement pointing away from the lens from the observer's standpoint. The displacement has components parallel to the source-lens relative motion, $\deltapar$, and perpendicular to it, $\deltaperp$, which can be expressed (e.g. \citealt{dominik00}) as

\begin{gather}
\deltapar=\frac{p}{\uz^2 + p^2 + 2}\thetae \nonumber \\
\deltaperp=\frac{\uz}{\uz^2 + p^2 + 2}\thetae \label{eq:deltacomp}\, ,
\end{gather}

\noindent
where $\uz$ is the impact parameter, or minimum source-lens angular separation, in units of $\thetae$. This occurs at time $\tz$, and

\begin{equation}\label{eq:p}
p\equiv p(t)=\frac{t-\tz}{\te}\, ,
\end{equation}

\noindent
where $t$ is the time, and $\te$ is the Einstein timescale, which is the time taken by the source to cross the Einstein ring radius, such that $\te=\thetae/\muls$, where $\muls$ is the source-lens relative motion. \Eq{eq:deltacomp} assumes a rectilinear uniform source-lens relative motion, and is independent of the observational point spread function (PSF). For a detailed discussion of the behaviour of the expressions in \Eq{eq:deltacomp}, see \cite{dominik00}. As the source moves relative to the lens, the components of the astrometric shift lead to a characteristic 1-dimensional (\Fig{fig:imbh_shift_time}) pattern, or, in 2 dimensions, to an elliptical motion of the source's centroid, as shown in \Fig{fig:imbh_shift_pos}. These ellipses have eccentricity $\epsilon=[2/(u_0^2 + 2)]^{1/2}$ \citep{dominik00}.

The photometric and astrometric effects behave differently at small and large separations. From Eqns. (\ref{eq:muu}) and (\ref{eq:deltau}), we see that for small values of $u$, the magnification becomes very large, while the astrometric signal decreases linearly with $u$. For large values of $u$, the photometric signal goes as $u^{-4}$, whereas the astrometric shift only decreases as $u^{-1}$. This means that the cross-section for astrometric events is significantly larger than for photometric ones, making them an interesting channel to detect events for which lenses are massive enough to cause a detectable signal.

\section{Measuring the mass of single objects with microlensing}\label{sec:massmeas}

Microlensing has been used to measure the mass of single stars, which is made possible when subtle second-order effects are detectable in the light curves \citep[e.g.][]{gould04b}. For instance, the lens-source parallax $\pils=\dl^{-1} - \ds^{-1}$ can be measured through light curve distortions, which means that the Einstein radius can then be constrained using \Eq{eq:thetae}. If the size of the source can also be constrained via additional second-order (``finite source size") effects \citep[e.g.][]{gould92}, then we can measure $\thetae$ and obtain a mass estimate for the lens \citep[e.g.][]{kains13a}. Recently, observations from space telescopes have been used to constrain the parallax in microlensing events \citep[e.g.][]{street16, zhu15}, with plans to do this more routinely for events of interest with the \textit{Spitzer Space Telescope} \citep[e.g.][]{udalski15, yee15}. 

In the case of lensing by an IMBH, however, finite source size and parallax effects will not usually be detected, because most events will only be detectable through astrometry, and not photometry, due to the much larger cross-section for astrometric events. Furthermore, for the kind of deep observing campaigns towards the Galactic Bulge that would be optimal for IMBH searches, the overwhelming majority of source stars will be main-sequence stars, which are too small to produce significant source-size effects.

Because the Einstein radius for an object scales with $\sqrt{M}$ (Eq. \ref{eq:thetae}), observations over many years are needed to detect signals from IMBH lensing that allow us to constrain the properties of the lens, whereas observations spanning months to a couple of years are usually sufficient for stellar-mass lenses. However, despite the extreme event timescales produced by IMBH lenses, they also lead to a much larger astrometric signal, making them easier to detect than for low-mass lenses.

If an astrometric signal is detected, the elliptical motion of the source's centroid can be used to measure $\thetae$, via \Eq{eq:deltacomp}. In the case of field lens objects, only analysis of second-order effects in the photometric event's light curve can then yield a constraint on $\dl$, in order to combine it with $\thetae$ to obtain a lens mass measurement. However, when $\dl$ is known, as is the case when considering IMBH lenses in the cores of globular clusters, the detection of the photometric event is not necessary. We can derive or assume a value $\ds$ (e.g. the distance to the Galactic Bulge for Bulge sources), so that the lens mass can be obtained from an astrometric detection only, through \Eq{eq:thetae}. To do this from the time-series astrometric measurements, we fit the elliptical trajectory due to lensing simultaneously with the source proper motion parameters. The lensing event can be characterised with the parameters $\tz$, $\te$, $\uz$, $\thetae$, as well as an inclination angle $\alpha$ of the lens-source motion, while four parameters are needed for the source proper motion: motions along the $x$ and $y$ axes, $\mu_x$ and $\mu_y$, as well as arbitrary reference points $x_0$ and $y_0$.

\section{Feasibility of an IMBH detection with microlensing}\label{sec:feasibility}

\subsection{Selection of cluster candidates}

Usually, campaigns focusing on stellar microlensing towards the Galactic Bulge (or other crowded regions) require careful estimates of the optical depth for both photometric and astrometric microlensing. In order to do this, one has to consider the entire populations of potential lens and source stars between the observer and the Galactic Bulge. When searching for IMBH in globular clusters, however, this is greatly simplified because the location of the IMBH is known, in so far as the distance to the cluster is known. Normally this is the case to within a precision of 0.5 kpc or better for Galactic clusters. 

In theory, an IMBH in a cluster can lens both stars within the cluster itself and background stars. In practice, however, it is clear from \Eq{eq:thetae} that the Einstein radius, and therefore the lensing cross-section, tends to 0 for $\dl \sim \ds$. Therefore the overwhelming probability for lensing comes from cases in which the source is a background star. For this reason, detections are only likely for clusters that lie in front of the Galactic Bulge, the Small Magellanic Cloud (SMC), or the Large Magellanic Cloud (LMC), where background star number densities are high enough that a lensing event is reasonably likely to occur. This limits the sample of clusters to be considered. Furthermore, simulations have shown that IMBHs are highly unlikely to exist in core-collapsed clusters \citep{baumgardt05}, which excludes a significant number of targets. We also rejected clusters in high-extinction areas, only selecting clusters with a horizontal branch (HB) brighter than 19 mag in $V$. Clusters with a fainter HB suffer from high extinction, meaning that background stars would also be highly extinguished, and detecting sufficient numbers of them with a good enough signal-to-noise ratio (SNR) would require very long exposure times. We excluded low-mass clusters such as Al 3 (BH 261), Djorg 2 (ESO456-SC38), and NGC 6540, and we rejected NGC 6809 because it is far away from the Bulge, and has a low density of background stars, dominated by inner halo stars. The final list of clusters fulfilling all criteria is given in \Tab{tab:gclist}. We note that among these, NGC 362 is possibly currently undergoing core collapse \citep[e.g.][]{dalessandro13}, but we include it because there is still some debate as to the dynamical status of this cluster \citep[e.g.][]{mclaughlin05}. We also note that \cite{leigh14} used $N$-body simulations to show that clusters containing stellar-mass black hole in binary systems were less likely to host IMBHs with masses higher than $\sim 10^3 \msun$. M~22 contains two known stellar-mass black holes \citep{strader12}, possibly meaning that any IMBH in this cluster is likely to have a mass lower than $\sim 10^3 \msun$. However, the half-mass relaxation time of M~22 is $\sim 2$ Gyr \citep{harris96}, which is long enough for a higher-mass IMBH to co-exist with stellar-mass black holes that might still remain \citep{heggie14}. Probing the existence of an IMBH in this cluster would therefore provide an excellent opportunity to test the simulation results of \cite{leigh14}.

Another consideration is that the optical depths for photometric and astrometric microlensing have different dependences on the lens and source distances. \cite{dominik00} showed that, with $x=\dl/\ds$, the optical depth for astrometric microlensing goes as $(1-x)^2$, while for photometric microlensing the dependence goes as $x(1-x)$. This means that while for photometric events, the lensing probability peaks for $x=0.5$, i.e. a lens located half-way between the observer and the source, for astrometric events, the probability is highest for lenses that are much closer to the observer than the source, i.e. $x\sim0$.

\begin{table*}
\begin{center}
  \begin{tabular}{ccccccccccccc}
  \hline
    ID			   &RA		&DEC     		& $m_{V, \rm HB}$ &$r_c$	& Dist 	&$s$			&$\muls$		\\
    			   &[J2000.0]	&[J2000.0]     	& [mag] 			&[arcmin]	& [kpc] 	&[arcsec$^{-2}$]	&[mas/yr]\\
  \hline  
   Bulge \\
  \hline
   NGC 6121 (M~4)	&16:23:35  &-26:31:33       &13.45	&1.16    	&2.2		&0.10$^a$		&$16.0^b$\\
   NGC 6304		&17:14:32  &-29:27:43	&16.25	&0.21	&5.9		&0.35$^c$		&$3.0^{d, *}$\\
   NGC 6528		   &18:04:50 &-30:03:23	&16.95	&0.13	&7.9		&3.2$^e$			&$1.3^{e}$\\
   NGC 6553	           &18:09:16 &-25:54:28	&16.60	&0.53	&6.0		&1.6$^f$			&$5.9^{f}$\\  
   NGC 6626 (M~28)      &18:24:33  &-24:52:11   &15.55	&0.24	&5.5   	&1.5$^\ddag$		&$4.9^{g}$\\
   NGC 6656 (M~22)	  &18:36:24 &-23:54:17	&14.15	&1.33	&3.2		&1.3$^h$			&$12.2^{h, \dagger}$\\    
\hline
   SMC \\
  \hline	
   NGC 104 (47 Tuc)	&00:24:06  &-72:04:53	&14.06	&0.36	&4.0		&0.02$^h$			&$4.9^{i}$\\
   NGC 362		&01:03:14  &-70:50:56	&15.44	&0.18	&8.6		&0.09$^h$			&$5.9^{h, \dagger}$\\
\hline \hline
  \end{tabular}
  \caption{Selected cluster targets. HB magnitudes, core radii, and distances are taken from the catalogue of \protect\cite{harris96}, except for the distance to NGC 104, which is from \protect\cite{mclaughlin06}. References for the number densities of background stars $s$ and proper motions relative to background Bulge/ SMC stars $\muls$ are (a)\protect\cite{bedin13}, (b)\protect\cite{dinescu99} (c)\protect\cite{sarajedini07}, (d)\protect\cite{dinescu03}, (e)\protect\cite{lagioia14}, (f)\protect\cite{zoccali01}, (g)\protect\cite{casetti13}, (h)\protect\cite{bellini14}, (i)\protect\cite{anderson03}. $^*$denotes an absolute proper motion measurement. $^\dagger$denotes a value of $\muls$ calculated from a proper motion catalogue rather than taken from a reference paper. $^\ddag$For NGC 6626, we estimated the stellar density based on its location along a straight line between NGC 6656 and NGC 6553. \protect\label{tab:gclist}}
  \end{center}
\end{table*}

\subsection{Event timescales}
For a stellar-mass lens event detected in an observing campaign lasting $\tobs$, generally, $\te \ll \tobs$, so that the photometric event can be observed from baseline to peak and back to baseline as long as the magnification of the source reaches above a threshold. On the other hand, the astrometric event unfolds much more slowly, with $\tast > \tobs$, and can usually not be observed in its entirety, except for low-mass lenses. Instead it is detected as long as the variation in centroid position over the time $\tobs$, $\delta_{\mathrm{obs}}$, is above a threshold $\deltat$, as discussed by \cite{dominik00}. Therefore the centroid shift itself, for the full event, might be larger than $\deltat$, but a short observing campaign may not be able to detect the event because variations are slow, resulting in $\delta_{\mathrm{obs}} < \deltat$.

For high-mass lenses such as IMBHs, $\tast \gg \tobs$ for typical observing campaigns. Furthermore, because the cross-section for detection of astrometric events is much larger, and an astrometric detection is sufficient to constrain the IMBH mass when $\dl$ is known, we will not consider the detection of the photometric event.

\begin{figure}
  \centering
  \includegraphics[width=8cm, angle=0]{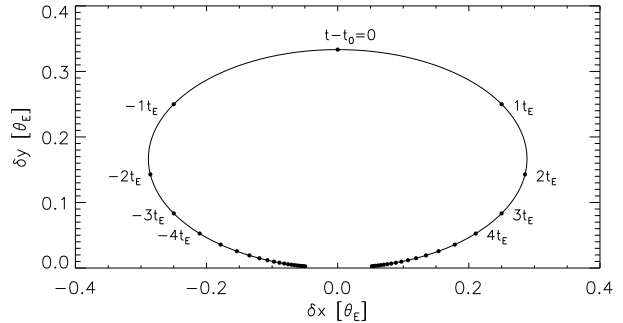}
  \caption{2-D astrometric motion due to a lensing event, as a function of time $t-\tz$ (in units of $\te$), shown here for $\uz=1$. The black filled circles are spaced equally by 1$\te$. The source position changes more rapidly as it nears its closest approach to the lens at $t=\tz$. \protect\label{fig:pos_te}}
\end{figure}

In addition to this, over large portions of the astrometric event, the centroid shift of the source caused by IMBH lensing will be linear and uniform in time, making it indistinguishable from the centroid shift due to proper motion of the source. Therefore only observations covering certain parts of the astrometric curve allow us to disentangle the real lensing effect from the proper motion. This corresponds to parts of the astrometric curve where the change in total displacement is not uniformly changing over $\tobs$, as shown in \Fig{fig:imbh_shift_time}, or for the 2-dimensional astrometric motion, as shown in \Fig{fig:imbh_shift_pos}. This 2-dimensional motion can allow us to detect curvature in the astrometric change even when the 1-dimensional displacement appears uniform. The 2-D astrometric motion (Fig. \ref{fig:imbh_shift_pos}) unfolds faster close to $t=\tz$ (\Fig{fig:pos_te}), so for a given $\tobs$, the fraction of the 2-D astrometric curve that is covered is larger when the position of the source is closer to $\uz$.

\subsection{Defining event rates}

We carried out simulations to determine the probability of an IMBH being detected unambiguously from an astrometric microlensing event in each cluster that passed our selection criteria. We considered 9 different IMBH masses, $\mbh/\msun=10^6$, $5\times 10^5, 10^5, 5\times 10 ^4, 10^4, 5 \times 10^3$, $10^3$, $5 \times 10^2$, and $10^2$. Although we considered masses up to $10^6\msun$ in order to investigate the full mass range of IMBHs, an IMBH with a mass larger than $\sim10^5\msun$ has a typical sphere of influence \citep{peebles72} of $\sim 0.1-5'$, which is comparable to, or larger than the core radius of globular clusters. Therefore, such IMBHs could also be detected by a number of other techniques, since their surface brightness profiles would not be well fitted by King models, for example.

For each cluster, $\dl$ is fixed by the distance to the cluster, and we assume $\ds$ is the distance to the Bulge or the SMC. The cluster distances we used are given in \Tab{tab:gclist}, and we assume distances to the Bulge and SMC of 8.5 kpc, which is a value within the range of different estimates in the literature \citep[e.g.][]{eisenhauer03, gillessen09, vanhollebeke09} and 61 kpc \citep{hilditch05}, respectively.

Our ability to detect a lensing event unambiguously depends on a number of factors, and the expected number of detections can be expressed as

\begin{equation}\label{eq:pdet}
\avn=\int_{-\infty}^{\infty}\int_{-\infty}^{\infty}s(x, y) \, \pdet (x, y) \,dx\, dy\, ,
\end{equation}

\noindent
where $s(x, y)$ is the number density of background stars at coordinates $(x, y)$, and $\pdet (x, y)$ is the probability for a single star that the event will be detected unambiguously, using criteria described in \Sec{sec:detectioncriteria}. The $(x, y)$ coordinate system is fixed relative to the IMBH, which may be assumed to be at the origin.

\subsubsection{Astrometric signals from binaries}\label{sec:binaries}

Although necessary, the detection of curvature in the astrometric curve of a background star is not sufficient to guarantee the detection of a lensing event. Indeed, with most stars being members of binary, or multiple systems, and the wide range of orbital separations and eccentricities of such systems \citep[e.g.][]{raghavan10}, many stars will exhibit astrometric signatures due to orbital motion around the binary system's centre of mass. Although some binaries are easily distinguished from single stars via their position on a colour-magnitude diagram (CMD), this is complicated in the Bulge by a number of factors such as the metallicity spread of stars, differential reddening, the range of distances due to the size of the Bulge, and some contamination from Disk stars. It is therefore useful to investigate the extent to which signals caused by orbital motion in a binary might mimic signals caused by the lensing of source stars by an IMBH along the line of sight.

Little is known about the frequency of binaries in the Galactic Bulge or the SMC, and the distribution of their orbital separations and eccentricities. In order to quantify how much of a confounding factor astrometric binaries can be in this study, we use the distribution for Disk stars of \cite{raghavan10}, who found a Normal period distribution with $\langle \log P \rangle=5.03$ and $\sigma_{\log P}=2.28$, for $P$ given in days. They also found an approximately uniform distribution of eccentricities between $e=0$ and $\sim 0.9$, except for systems with $P < 12$d, which are circularised ($e=0$). We assumed the scenario for Bulge stars that would produce the largest astrometric signal, which corresponds to equal-mass binaries and a total mass of $1.9\msun$, the largest possible total binary mass in the Bulge \citep{calamida15, duquennoy91}. We also used uniform distributions for the orientation of the system with respect to the $x$, $y$, and $z$ axes.

\begin{figure}
  \centering
  \includegraphics[width=8cm, angle=0]{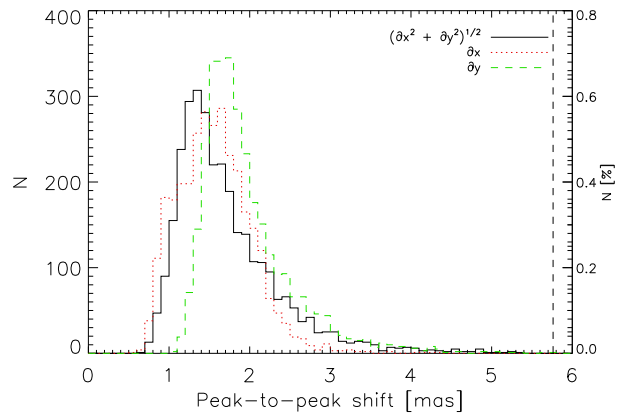}
  \caption{Histogram of the peak-to-peak astrometric shift detected from astrometric binaries in the Bulge, for equal-mass binaries of total mass $1.9\msun$. The vertical dashed line indicates the highest shift that can be caused by a binary companion in the Bulge. \protect\label{fig:histbinaries}}
\end{figure}

For Bulge stars, we find that shifts with a peak-to-peak amplitude of up to $\sim 6$ mas can be caused by astrometric binaries, with most amplitudes between $\sim$1 and 3 mas, as shown in \Fig{fig:histbinaries}. It is therefore clear that we must account for the possibility of a binary orbit explaining an astrometric signal if the amplitude is smaller than $\sim 6$ mas. To do this, we fit a binary orbit model to each set of our simulated astrometric data, and consider an astrometric lensing event as unambiguously detected only if no competing binary model can be fitted. For SMC stars, no binary produces a peak-to-peak astrometric shift above 0.4 mas. Since this is the best astrometric precision that can be achieved for bright sources (see Eq. \ref{eq:precision} below), we can therefore safely assume that any event that shows peak-to-peak shifts larger than that is caused by lensing.

\subsubsection{Orbits of cluster members}\label{sec:clustmembers}

In theory, stars that are cluster members could also have non-uniform astrometric curves, due to their orbits inside of the cluster. These could then potentially be confused for background stars being lensed by an IMBH in the cluster. However, we find that the number of such stars is negligible, due to the extremely long orbits involved. Furthermore, in the vast majority of cases, the combination of proper motions and the CMD allows us to determine cluster membership, even in cases where the cluster's bulk motion relative to the background stars is small \citep[e.g.][]{lagioia14}. Finally, we have conducted simulations to estimate the number of astrometric lensing curves that can be mimicked by circular orbits of cluster members, and found this to be within the error bars of our results (see \Sec{sec:discussion}).

\subsection{Estimates of background star number densities and $\muls$}

In order to estimate $\avn$ for each cluster, we must first estimate the number density $s$ of background stars. The number density we calculate here is for stars with $m_{\rm F814W} \leq 26$ mag, which corresponds to the faintest Galactic Bulge stars that are detected with WFC3/UVIS in 15-minute observations with a SNR better than 2.5.

We used proper motion catalogues, when available, in order to separate cluster stars from the Bulge and Disk populations; references for the catalogues we used are given in \Tab{tab:gclist}. When possible, we estimated the completeness of the proper motion catalogues by comparing the corresponding photometric catalogues to the ones from the \textit{Hubble Space Telescope} (HST)  \textit{Advanced Camera for Surveys} (ACS) Survey of Galactic Globular Clusters \citep{sarajedini07}. We then compared the bright end ($19 < \m814 < 20.5$) of the resulting number density distribution of background stars to the distribution of Bulge stars of \cite{calamida15} in the Sagittarius Window Eclipsing Extrasolar Planet Search (SWEEPS) field ($l=1.25^{\circ}, b=-2.65^{\circ}$). We used the bright end of the distribution because this is where the photometry is not affected by completeness issues. We calculated the number density in the SWEEPS field by using the mass function found by \cite{calamida15} to evaluate star counts down to $m_{\rm F814W}\sim 26$ mag, and found an average SWEEPS field density of 6.5 stars/ arcsec$^2$. Finally, we used this to derive a scaling factor for number densities along the line of sight to the globular clusters in our sample. For the clusters toward the SMC, we used the same process, but with the SMC luminosity function of \cite{kalirai13} instead of the Bulge distribution, and found an average SMC number density of 0.04 stars/ arcsec$^2$.

For the source-lens relative motion $\muls$, required as an ingredient of our simulations, we used values from the literature from proper motion studies, taking the bulk relative proper motion of the cluster as a proxy for $\muls$. For NGC~362, we derived a value of $\muls$ from the proper motion catalogue of \cite{bellini14} by calculating the median proper motion of cluster and Bulge stars; we also did this calculation for NGC~6656 and NGC~104, to check that we obtained values consistent with those we adopted from the literature. The resulting values of $\muls$ are listed, and relevant references are given, in \Tab{tab:gclist}.

We adopt a scatter in the value of $\muls$ for the background stars, using the dispersion in proper motion for Bulge stars of 2.6 mas/yr along the directions of both velocity components. This value is in line with that of \cite{clarkson08}, but we adopt the same value in both directions for simplicity. For SMC stars we use a scatter of 0.3 mas/yr for both directions, in agreement with the findings of \cite{vieira10}.

\subsection{Simulations}\label{sec:simulations}

We adopt a Monte Carlo approach to evaluating the detection probabilities $\pdet(x,y)$ over a grid in $(x,y)$ for each IMBH mass. To keep the number of simulations reasonable as we evaluate $\pdet(x,y)$ further away from the IMBH, we chose to perform a constant number of simulations, 1000, in rings of equal widths around the origin. We chose a ring width of 0.2$\thetae$.

For each simulation, we draw the position of a background star from a uniform distribution over the area of the current grid element and assume that this is the source position at $t=0$. Without any loss of generality, we assume that our $(x,y)$ coordinate system is aligned such that the bulk cluster proper motion relative to the background stars makes the background stars appear to move along the x-axis in the positive direction (since our IMBH is fixed at the origin). Considering also that the background stars exhibit a velocity dispersion, we therefore draw the source-lens relative proper motion from a two-dimensional Gaussian distribution with means in the $x-$ and $y-$directions of $\muls$ and zero, respectively, and $\sigma$ in both directions equal to the dispersion in the Bulge or SMC proper motions as appropriate.

In the HST data archive, there are typically of the order of 50 images for a cluster, spaced out over 20 years. Hence, for each simulation we adopt a time baseline of $\tobs=20$ years with the first and last images obtained at $t=0$ and $t=20$ years, respectively. The epochs of the remaining 48 images are drawn from a uniform distribution on the range [0, $\tobs$]. The astrometric motion curve of the background star is then generated using these epochs taking into account the source-lens relative proper motion, its position at $t=0$, and the lensing effect of the IMBH.

To simulate measurement noise in the astrometric motion curve of the background star, we assign the star a random magnitude, drawn from the Bulge or SMC luminosity functions of \cite{calamida15} or \cite{kalirai13}, respectively. We calculated the SNR in a 15-minute exposure with WFC3/UVIS for different magnitudes between $\m814=18$ and 26 mag, with a bin size of 0.2 mag, using the HST exposure time calculator. This allowed us to estimate the SNR for each background star, and thereby, the astrometric measurement precision $z$, through the expression \citep{kuijken02},

\begin{equation}\label{eq:precision}
z=0.7\frac{\mathrm{FWHM}}{\mathrm{SNR}\times \sqrt{N_e}}\, ,
\end{equation}

\noindent
where FWHM is the full-width half-maximum of the star's PSF, and $N_e$ is the number of images per epoch. This level of precision has been routinely achieved by several projects using HST observations for high-precision astrometric measurements \citep[e.g.][]{bedin13, bellini14, vandermarel14}. For simplicity, we use the pixel scale of WFC3/UVIS of 40 mas/ pixel for all simulated observations, and we use a conservative estimate of $N_e=4$. Each astrometric measurement of the background star is perturbed by a random number drawn from a Gaussian distribution with zero mean and $\sigma=z$, thereby generating our simulated noisy astrometric motion curve from HST observations.

\subsection{Detection criteria}\label{sec:detectioncriteria}

For our detection criteria, we need a statistic that will allow us to discriminate between astrometric motion models with different numbers of parameters (i.e. rectilinear uniform motion, astrometric microlensing, and orbital motion). We use the Bayesian Information Criterion (BIC; \citealt{schwarz78}) derived by approximating the posterior probability of each model. It is valid for model parameters estimated by maximum likelihood. For model parameters with a uniform prior, it is given by the expression

\begin{equation}\label{eq:bic}
\mathrm{BIC}=-2\ln(\mathcal{L}) + \np\ln(\nd)\, - \np\ln(2\pi)
\end{equation}
\noindent
where $\mathcal{L}$ is the maximum-likelihood statistic, $\nd$ is the number of data points, and $\np$ is the number of model parameters. The use of an information criterion for discriminating between models in our simulations is particularly appropriate since we have generated the astrometric measurements for each background star from independent Gaussian distributions for which the sigma values are known exactly. In this case, the BIC further reduces to

\begin{equation}\label{eq:bic}
\mathrm{BIC}=\chi^2 + \np\ln(\nd)\, - \np\ln(2\pi) + K
\end{equation}
\noindent
where $\chi^2$ is the chi-squared statistic, and K is a constant term that can be ignored for model selection purposes. The ratio of the posterior probabilities $P(A)$ and $P(B)$ of two models may be calculated from the BIC via:

\begin{equation}\label{eq:bicprob}
\frac{P(A)}{P(B)}=\exp(0.5(\mathrm{\dbic)})\, ,
\end{equation}
\noindent
where $\dbic=\mathrm{BIC_B}-\mathrm{BIC_A}$. We choose to adopt the threshold corresponding to a relative probability $P(A)/P(B)=100$, i.e. $\dbict$=9.21.

For each noisy astrometric motion curve that we generated in the simulations, we first fit a rectilinear uniform proper motion model (4 parameters) to the data and we calculate a corresponding BIC value which we denote as $\biclin$. We then perform a slew of tests which must be satisfied in order for a successful detection of astrometric microlensing event to be declared:

\begin{enumerate}
\item We compute the peak-to-peak amplitude $\dobs$ of the residuals to the fit. If this is above 2z, then we proceed to step (ii).
\item We fit the 9-parameter astrometric lensing model to the data (see Section 3) and calculate the corresponding $\biclens$.
\item We check that the mass of the IMBH lens is recovered correctly to within a factor of ten, and if so, then we proceed to step (iv).
\item We verify that the astrometric lensing model is favoured over the rectilinear uniform proper motion model above our threshold, i.e. $\biclin - \biclens \ge \dbict$. If so, then we proceed to step (v).
\item If $\dobs \ge $ 6 mas, then we know that the astrometric signal cannot be explained by a binary orbit, and we declare successful detection of astrometric microlensing. Otherwise, we proceed to step (vi)
\item We fit the 11-parameter orbital motion model to the data (four parameters for the rectilinear uniform source proper motion, and seven parameters for binary orbital motion, see Section 4.3.1) and calculate the corresponding $\bicbin$.
\item We verify that the astrometric lensing model is favoured over the orbital motion model above our threshold, i.e. $\bicbin - \biclens \ge \dbict$. If so, then we declare successful detection of astrometric microlensing. Otherwise, we finish.
\end{enumerate}

\begin{figure}
  \centering
  \includegraphics[width=8cm, angle=0]{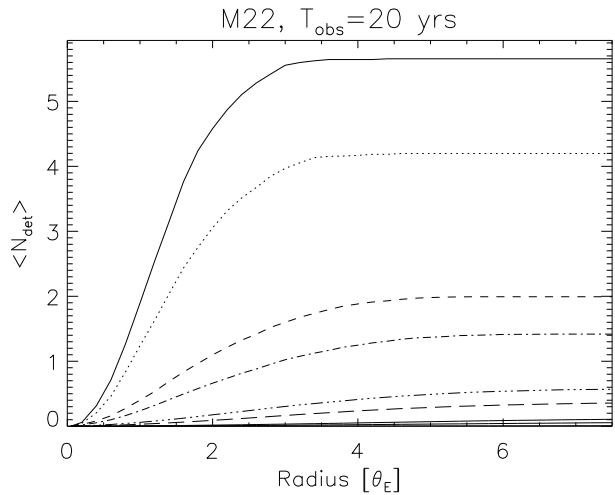}
  \caption{Cumulative expected number of detections of IMBH lensing events, as a function of the distance from the lens, for each of the nine IMBH masses considered, with $\avn$ increasing with mass. \protect\label{fig:prob_nthe}}
\end{figure}

The computation of $\pdet(x,y)$ for each grid element is then trivial as the ratio of the number of successful detections to the number of simulations performed. The detection probabilities tend to zero as the distance from the IMBH increases due to lack of curvature and decreasing peak-to-peak signal in the astrometric motion curves over the observational baseline. In \Fig{fig:prob_nthe}, we plot $\avn$ evaluated via equation 6 as a function of integration radius from the IMBH. We can stop the integration when asymptotic limits for $\avn$ are reached, for example at $\sim6 \, \thetae$ for an IMBH in M~22 and $\tobs=20$ years.

\section{Discussion}\label{sec:discussion}

Results from our simulations are given in \Tab{tab:gcndet}. 

From this, we see that $\avn$ is significant for most of the selected Bulge clusters for IMBH masses above $\sim 10^4 \msun$, and for some, down to masses of $\sim 10^3\msun$. Unsurprisingly, the most promising clusters for such detections are four of the five clusters in our sample closest to the Solar System, with M~22, NGC~6553, NGC~6121, and NGC 6626 having the largest number of expected events. For clusters toward the SMC, low background star densities make lensing probabilities, and $\avn$, very low.

\begin{figure*}
  \centering
  \includegraphics[width=8cm, angle=0]{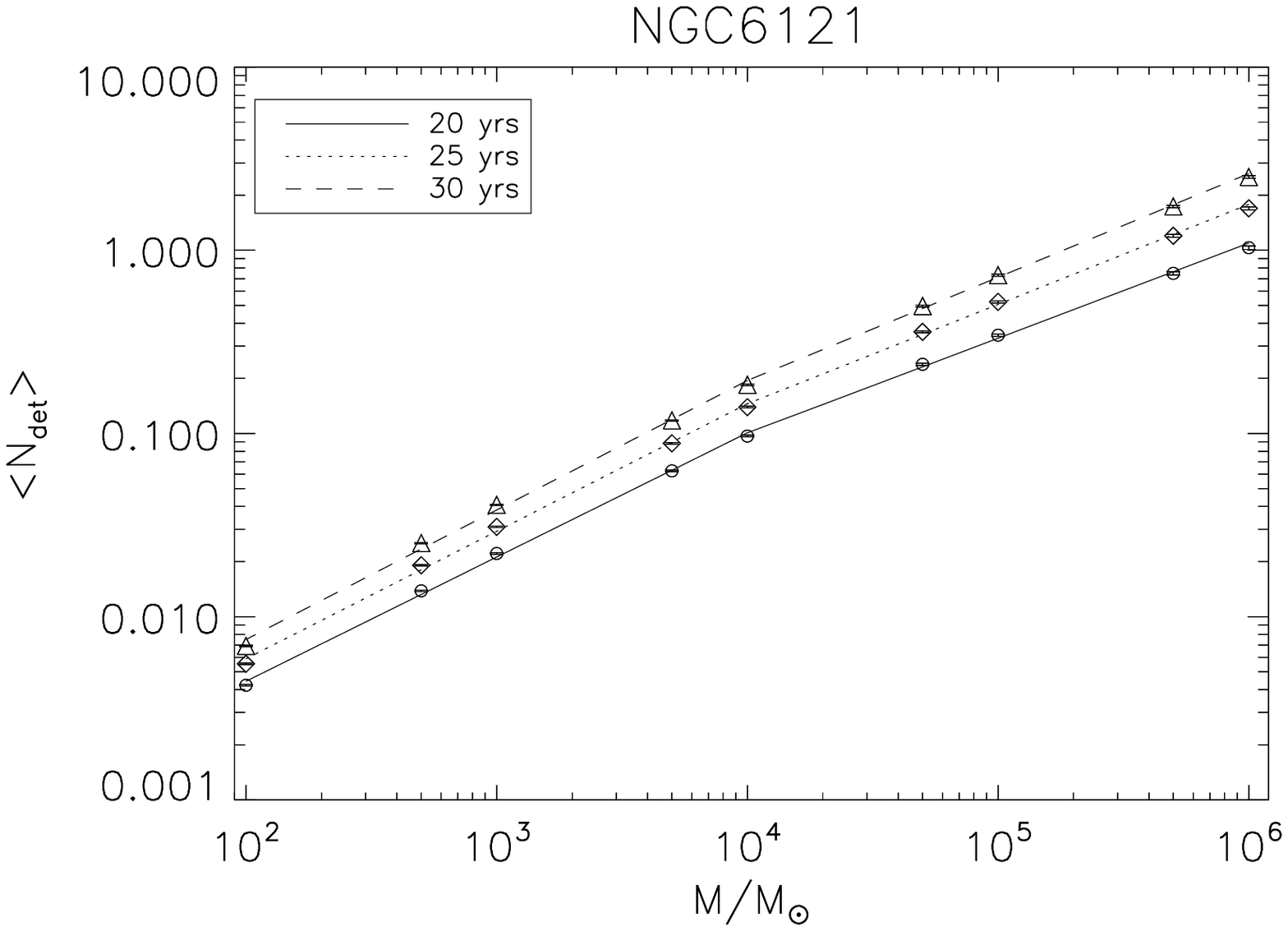}
  \includegraphics[width=8cm, angle=0]{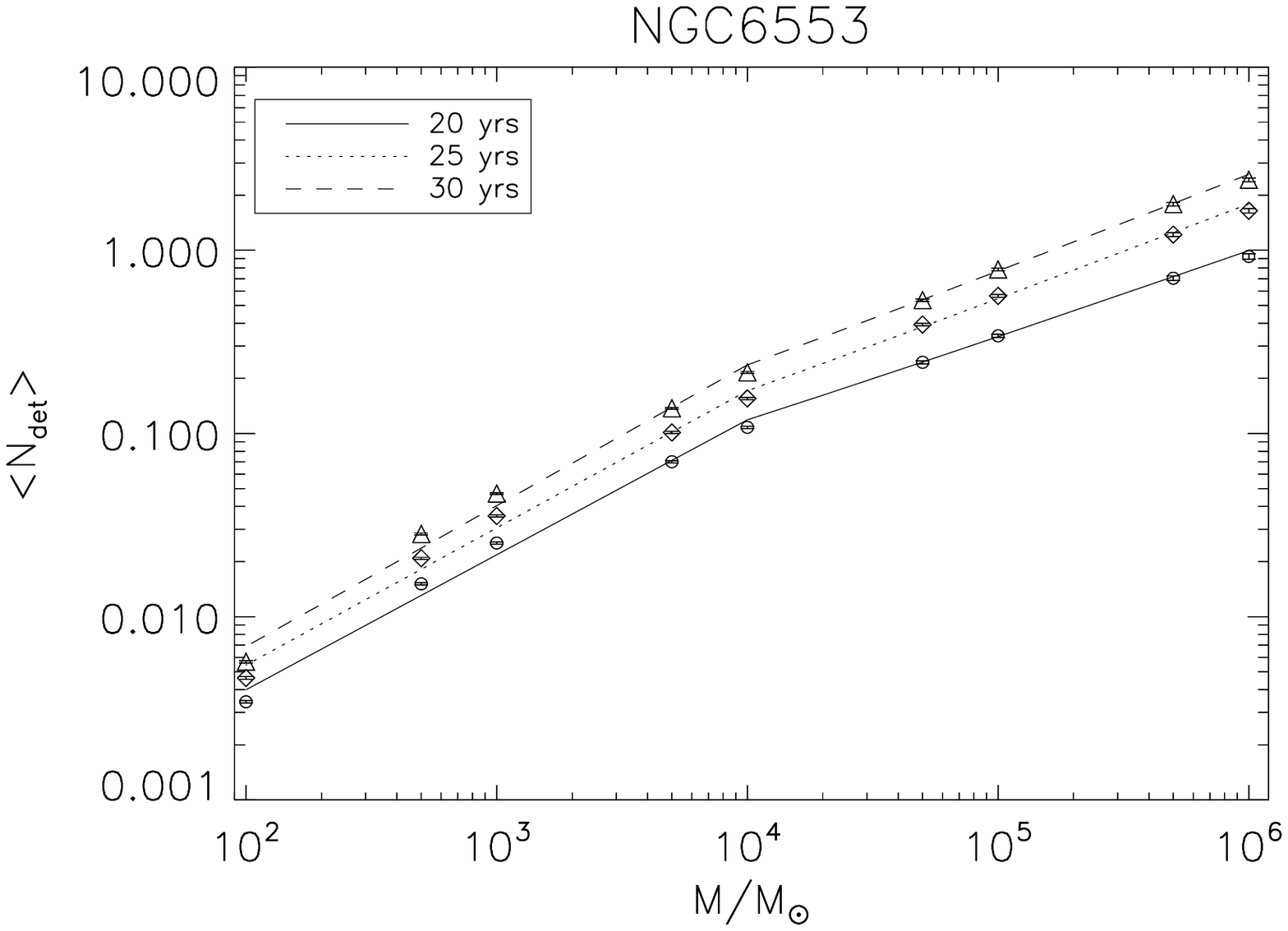}
  \includegraphics[width=8cm, angle=0]{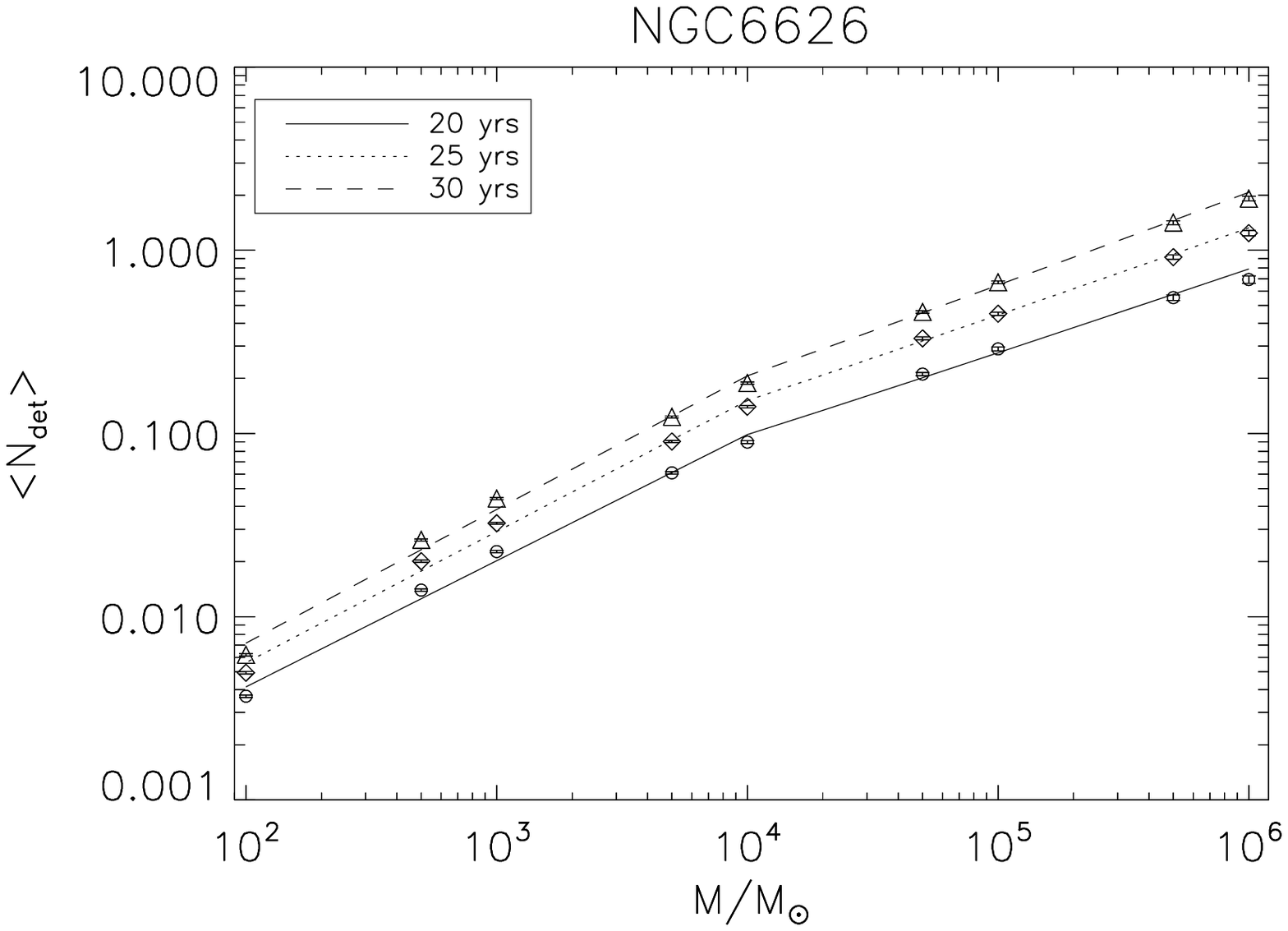}
  \includegraphics[width=8cm, angle=0]{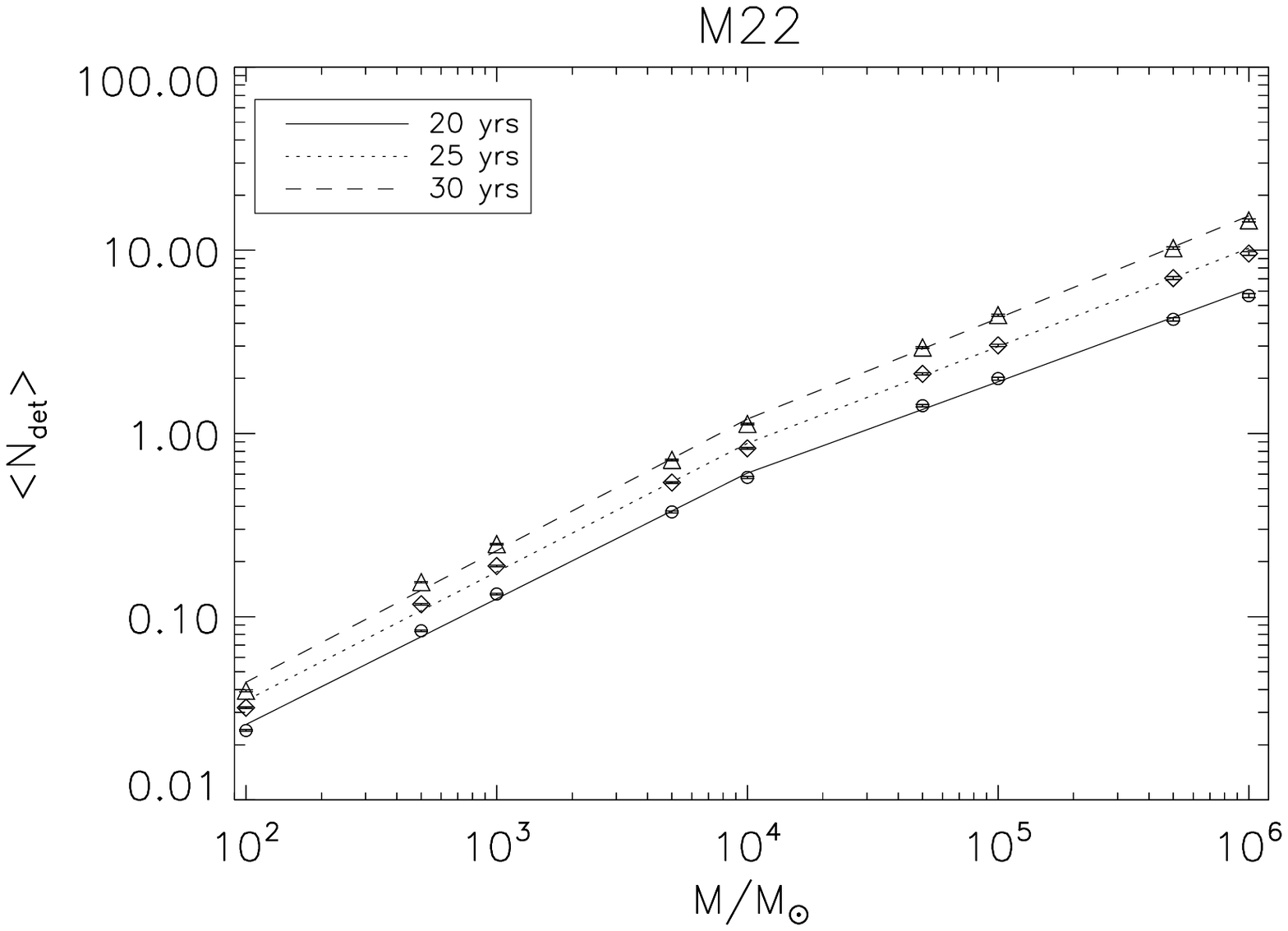}
  \caption{\small{The expected number of detected events as a function of black hole mass. The detection rates from simulations are plotted as triangles, diamonds, and open circles, along with a power-law fit, for baselines of 20, 25, and 30 years, respectively.} \label{fig:mass_ndet}}
\end{figure*}

We plot $\avn$ as a function of lens mass for baselines of 20, 25, and 30 years for these four clusters in \Fig{fig:mass_ndet}. With a time baseline of 20 years, $\avn$ in M~22 is large for $M > 3\times10^4\msun$, and declines with mass down to $\avn=0.1$ at $M=\sim 10^3 \msun$. This makes it the best candidate for further analysis. The next best candidate is NGC~6553, with $\avn = 0.93$, 0.70, and 0.34 for $M=10^6$, $5\times 10^5$ and $10^5 \msun$ respectively, and $\avn=0.11$ at $M=10^4\msun$. Thanks to the fast motion of M~4 relative to the Bulge, and despite low stellar densities, the expected numbers of events are slightly higher than NGC~6553 for a high-mass IMBH, with $\avn = 1.03$ and 0.75 for $M=10^6$ and $5\times10^5 \msun$ respectively.

For the Bulge clusters, we fit a power law for the mass dependence of $\avn$, of the form

\begin{equation}\label{eq:powerlaw}
\avn=a \left(\frac{M}{M_f}\right)^{b} ,
\end{equation}
\noindent
where $a$ and $b$ are the fitted power-law parameters, and $M_f$ is an arbitrary fiducial mass. We find that the mass dependence is best fitted with two mass regimes, and therefore we fit two power laws for masses below and above $10^4\msun$. The coefficients $a$ and $b$ for both regimes are given in \Tab{tab:powerlaw}.

\begin{table}
\begin{center}
  \begin{tabular}{lllll}
\hline
    	 	&$a_{\mathrm{LM}}$		&$b_{\mathrm{LM}}$		&$a_{\mathrm{HM}}$		&$b_{\mathrm{HM}}$	\\
\hline
   $\tobs=20$ years \\
  \hline	   
NGC 6121 (M 4) & 0.10(1) & 0.64(7) & 0.33(1) & 0.50(1) \\
NGC 6304 & 0.01(1) & 0.59(53) & 0.03(1) & 0.37(4) \\
NGC 6528 & 0.03(1) & 0.69(28) & 0.09(1) & 0.47(2) \\
NGC 6553 & 0.11(1) & 0.65(6) & 0.33(1) & 0.47(1) \\
NGC 6626 (M 28) & 0.09(1) & 0.61(7) & 0.27(1) & 0.44(1) \\
NGC 6656 (M 22) & 0.58(1) & 0.64(1) & 1.90(1) & 0.51(1) \\
\hline
   $\tobs=25$ years \\
\hline	   
NGC 6121 (M 4) & 0.14(1) & 0.66(5) & 0.51(1) & 0.53(1) \\
NGC 6304 & 0.02(1) & 0.63(36) & 0.05(1) & 0.42(3) \\
NGC 6528 & 0.04(1) & 0.72(20) & 0.14(1) & 0.50(1) \\
NGC 6553 & 0.16(1) & 0.65(4) & 0.54(1) & 0.51(1) \\
NGC 6626 (M 28) & 0.14(1) & 0.65(5) & 0.44(1) & 0.47(1) \\
NGC 6656 (M 22) & 0.84(1) & 0.65(1) & 2.94(2) & 0.54(1) \\
\hline
   $\tobs=30$ years \\
\hline
NGC 6121 (M 4) & 0.18(1) & 0.66(4) & 0.72(1) & 0.55(1) \\
NGC 6304 & 0.03(1) & 0.64(27) & 0.08(1) & 0.48(2) \\
NGC 6528 & 0.06(1) & 0.75(15) & 0.20(1) & 0.53(1) \\
NGC 6553 & 0.22(1) & 0.67(3) & 0.76(1) & 0.53(1) \\
NGC 6626 (M 28) & 0.19(1) & 0.64(4) & 0.64(1) & 0.50(1) \\
NGC 6656 (M 22) & 1.13(1) & 0.67(1) & 4.22(2) & 0.56(1) \\
\hline \hline
  \end{tabular}
  \caption{Power-law coefficients (see Eq. \ref{eq:powerlaw}), for the two mass regimes (subscripts LM and HM denote the low- and high-mass regimes, respectively, with the limiting mass between the two regimes set at $10^4\msun$), for the Bulge clusters, and $\tobs=20$ years. We used $M_f=10^4\msun$ and $10^5 \msun$ for the low- and high-mass regimes, respectively. \protect\label{tab:powerlaw}}
  \end{center}
\end{table}

It is also useful to turn values of $\avn$ into probabilities that are easier to interpret. Under the assumption that an IMBH exists in the cluster, then the discrete Poisson distribution is appropriate for the number of stars observed to feature a detectable astrometric signal caused by an IMBH. We know that the detectable astrometric events occur with an expected average rate of $\avn$ calculated from our simulations, and we can therefore express the probability of $n$ events being detected as

\begin{equation}\label{eq:probpoisson1}
P(n)=\frac{\avn^n}{n!}e^{-\avn}\, ,
\end{equation}
\noindent
from which we can express the probability of at least one event being detected as 

\begin{equation}\label{eq:probpoisson1}
P(n>0)=1- e^{-\avn}\, .
\end{equation}
\noindent

In \Tab{tab:gcprob}, we give $P(n>0)$ for each IMBH mass, cluster, and time baseline, calculated from the values of $\avn$ listed in \Tab{tab:gcndet}.

\begin{table*}
\begin{center}
  \begin{tabular}{ccccccccccc}
  \hline
$\tobs=20$ years  				&&&&&$M/\msun$ \\
  \hline
	&$10^6$	&$5\times10^5$	&$10^5$	&$5\times10^4$	&$10^4$	&$5\times10^3$     	&$10^3$		&$5\times10^2$ 		&$10^2$	\\
\hline
\hline
   Bulge \\
  \hline	   
NGC 6121 (M 4) & 1.03(2) & 0.75(2) & 0.34(1) & 0.24(0) & 0.10(0) & 0.06(0) & 0.02(0) & 0.01(0) &0.004(0) \\
NGC 6304 & 0.07(0) & 0.06(0) & 0.03(0) & 0.03(0) & 0.01(0) &0.008(0) &0.003(0) &0.002(0) &0 \\
NGC 6528 & 0.26(1) & 0.18(1) & 0.09(0) & 0.06(0) & 0.03(0) & 0.02(0) &0.005(0) &0.003(0) &0 \\
NGC 6553 & 0.93(3) & 0.70(2) & 0.34(1) & 0.24(0) & 0.11(0) & 0.07(0) & 0.03(0) & 0.02(0) &0.003(0) \\
NGC 6626 (M 28) & 0.69(3) & 0.55(2) & 0.29(1) & 0.21(0) & 0.09(0) & 0.06(0) & 0.02(0) & 0.01(0) &0.004(0) \\
NGC 6656 (M 22) & 5.66(15) & 4.20(10) & 1.99(3) & 1.42(2) & 0.58(1) & 0.37(0) & 0.13(0) & 0.08(0) & 0.02(0) \\
\hline
   SMC \\
  \hline	   
NGC 104 (47 Tuc) &0 &0 &0 &0 &0 &0 &0 &0 &0 \\
NGC 362 &0 &0 &0.002(0) &0.002(0) &0.002(0) &0.002(0) &0.001(0) &0 &0 \\
\hline \hline
  \hline
$\tobs=25$ years  				&&&&&$M/\msun$ \\
  \hline
	&$10^6$	&$5\times10^5$	&$10^5$	&$5\times10^4$	&$10^4$	&$5\times10^3$     	&$10^3$		&$5\times10^2$ 		&$10^2$	\\
\hline
   Bulge \\
  \hline	   
NGC 6121 (M 4) & 1.69(3) & 1.20(2) & 0.52(1) & 0.36(0) & 0.14(0) & 0.09(0) & 0.03(0) & 0.02(0) &0.006(0) \\
NGC 6304 & 0.14(1) & 0.11(0) & 0.06(0) & 0.04(0) & 0.02(0) & 0.01(0) &0.004(0) &0.003(0) &0.001(0) \\
NGC 6528 & 0.43(1) & 0.31(1) & 0.14(0) & 0.10(0) & 0.04(0) & 0.02(0) &0.008(0) &0.004(0) &0 \\
NGC 6553 & 1.64(5) & 1.22(3) & 0.56(1) & 0.39(1) & 0.16(0) & 0.10(0) & 0.04(0) & 0.02(0) &0.005(0) \\
NGC 6626 (M 28) & 1.24(4) & 0.92(3) & 0.45(1) & 0.33(1) & 0.14(0) & 0.09(0) & 0.03(0) & 0.02(0) &0.005(0) \\
NGC 6656 (M 22) & 9.60(21) & 7.05(13) & 3.03(4) & 2.12(3) & 0.83(1) & 0.54(1) & 0.19(0) & 0.12(0) & 0.03(0) \\
\hline
   SMC \\
  \hline	   
NGC 104 (47 Tuc) &0 &0 &0 &0.001(0) &0.001(0) &0.001(0) &0 &0 &0 \\
NGC 362 &0 &0.002(0) &0.006(0) &0.006(0) &0.005(0) &0.004(0) &0.001(0) &0.001(0) &0 \\
\hline \hline
  \hline
$\tobs=30$ years  				&&&&&$M/\msun$ \\
  \hline
	&$10^6$	&$5\times10^5$	&$10^5$	&$5\times10^4$	&$10^4$	&$5\times10^3$     	&$10^3$		&$5\times10^2$ 		&$10^2$	\\
\hline
   Bulge \\
  \hline	   
NGC 6121 (M 4) & 2.51(4) & 1.73(3) & 0.73(1) & 0.50(1) & 0.18(0) & 0.12(0) & 0.04(0) & 0.03(0) &0.007(0) \\
NGC 6304 & 0.25(1) & 0.19(1) & 0.09(0) & 0.06(0) & 0.02(0) & 0.02(0) &0.006(0) &0.004(0) &0.001(0) \\
NGC 6528 & 0.65(2) & 0.48(1) & 0.20(0) & 0.14(0) & 0.05(0) & 0.03(0) & 0.01(0) &0.005(0) &0 \\
NGC 6553 & 2.43(6) & 1.79(4) & 0.79(1) & 0.53(1) & 0.22(0) & 0.14(0) & 0.05(0) & 0.03(0) &0.006(0) \\
NGC 6626 (M 28) & 1.91(6) & 1.42(3) & 0.67(1) & 0.46(1) & 0.19(0) & 0.12(0) & 0.04(0) & 0.03(0) &0.006(0) \\
NGC 6656 (M 22) &14.55(27) &10.28(17) & 4.42(6) & 2.95(3) & 1.13(1) & 0.72(1) & 0.25(0) & 0.15(0) & 0.04(0) \\
\hline
   SMC \\
  \hline	   
NGC 104 (47 Tuc) &0 &0 &0.001(0) &0.002(0) &0.001(0) &0.001(0) &0 &0 &0 \\
NGC 362 &0.004(0) &0.009(0) & 0.01(0) & 0.01(0) &0.008(0) &0.005(0) &0.002(0) &0.001(0) &0 \\
\hline \hline
  \end{tabular}
  \caption{Expected number of detected events for each of the 9 IMBH masses considered, for $\tobs=20$, 25, and 30 years. 1-$\sigma$ error bars on the last decimal place are given in parentheses. An error bar of 0 indicates that the Poisson error is smaller than the precision quoted. \protect\label{tab:gcndet}}
  \end{center}
\end{table*}

\begin{table*}
\begin{center}
  \begin{tabular}{ccccccccccc}
  \hline
$\tobs=20$ years  				&&&&&$M/\msun$ \\
  \hline
	&$10^6$	&$5\times10^5$	&$10^5$	&$5\times10^4$	&$10^4$	&$5\times10^3$     	&$10^3$		&$5\times10^2$ 		&$10^2$	\\
\hline
\hline
   Bulge \\
  \hline	   
NGC 6121 (M 4) & 0.64(1) & 0.53(1) & 0.29(0) & 0.21(0) & 0.09(0) & 0.06(0) & 0.02(0) & 0.01(0) &0.004(0) \\
NGC 6304 & 0.07(0) & 0.06(0) & 0.03(0) & 0.02(0) & 0.01(0) &0.008(0) &0.003(0) &0.002(0) &0 \\
NGC 6528 & 0.23(1) & 0.16(1) & 0.09(0) & 0.06(0) & 0.03(0) & 0.02(0) &0.005(0) &0.003(0) &0 \\
NGC 6553 & 0.60(1) & 0.51(1) & 0.29(1) & 0.22(0) & 0.10(0) & 0.07(0) & 0.02(0) & 0.02(0) &0.003(0) \\
NGC 6626 (M 28) & 0.50(2) & 0.42(1) & 0.25(1) & 0.19(0) & 0.09(0) & 0.06(0) & 0.02(0) & 0.01(0) &0.004(0) \\
NGC 6656 (M 22) & 1.00(0) & 0.98(0) & 0.86(0) & 0.76(1) & 0.44(0) & 0.31(0) & 0.12(0) & 0.08(0) & 0.02(0) \\\hline
   SMC \\
  \hline	   
NGC 104 (47 Tuc) &0 &0 &0 &0 &0 &0 &0 &0 &0 \\
NGC 362 &0 &0 &0.002(0) &0.002(0) &0.002(0) &0.002(0) &0.001(0) &0 &0 \\
\hline \hline
  \hline
$\tobs=25$ years  				&&&&&$M/\msun$ \\
  \hline
	&$10^6$	&$5\times10^5$	&$10^5$	&$5\times10^4$	&$10^4$	&$5\times10^3$     	&$10^3$		&$5\times10^2$ 		&$10^2$	\\
\hline
   Bulge \\
  \hline	   
NGC 6121 (M 4) & 0.82(1) & 0.70(1) & 0.41(0) & 0.30(0) & 0.13(0) & 0.08(0) & 0.03(0) & 0.02(0) &0.006(0) \\
NGC 6304 & 0.13(1) & 0.10(0) & 0.06(0) & 0.04(0) & 0.02(0) & 0.01(0) &0.004(0) &0.003(0) &0.001(0) \\
NGC 6528 & 0.35(1) & 0.27(1) & 0.13(0) & 0.10(0) & 0.04(0) & 0.02(0) &0.008(0) &0.004(0) &0 \\
NGC 6553 & 0.81(1) & 0.70(1) & 0.43(1) & 0.33(0) & 0.14(0) & 0.10(0) & 0.03(0) & 0.02(0) &0.005(0) \\
NGC 6626 (M 28) & 0.71(1) & 0.60(1) & 0.36(1) & 0.28(0) & 0.13(0) & 0.09(0) & 0.03(0) & 0.02(0) &0.005(0) \\
NGC 6656 (M 22) & 1.00(0) & 1.00(0) & 0.95(0) & 0.88(0) & 0.56(0) & 0.42(0) & 0.17(0) & 0.11(0) & 0.03(0) \\
\hline
   SMC \\
  \hline	   
NGC 104 (47 Tuc) &0 &0 &0 &0.001(0) &0.001(0) &0.001(0) &0 &0 &0 \\
NGC 362 &0 &0.002(0) &0.006(0) &0.006(0) &0.005(0) &0.004(0) &0.001(0) &0.001(0) &0 \\
\hline \hline
  \hline
$\tobs=30$ years  				&&&&&$M/\msun$ \\
  \hline
	&$10^6$	&$5\times10^5$	&$10^5$	&$5\times10^4$	&$10^4$	&$5\times10^3$     	&$10^3$		&$5\times10^2$ 		&$10^2$	\\
\hline
   Bulge \\
  \hline	   
NGC 6121 (M 4) & 0.92(0) & 0.82(0) & 0.52(0) & 0.39(0) & 0.17(0) & 0.11(0) & 0.04(0) & 0.02(0) &0.007(0) \\
NGC 6304 & 0.22(1) & 0.17(0) & 0.08(0) & 0.06(0) & 0.02(0) & 0.02(0) &0.006(0) &0.004(0) &0.001(0) \\
NGC 6528 & 0.48(1) & 0.38(1) & 0.18(0) & 0.13(0) & 0.05(0) & 0.03(0) & 0.01(0) &0.005(0) &0 \\
NGC 6553 & 0.91(1) & 0.83(1) & 0.54(1) & 0.41(0) & 0.19(0) & 0.13(0) & 0.05(0) & 0.03(0) &0.006(0) \\
NGC 6626 (M 28) & 0.85(1) & 0.76(1) & 0.49(1) & 0.37(0) & 0.17(0) & 0.12(0) & 0.04(0) & 0.03(0) &0.006(0) \\
NGC 6656 (M 22) & 1.00(0) & 1.00(0) & 0.99(0) & 0.95(0) & 0.68(0) & 0.51(0) & 0.22(0) & 0.14(0) & 0.04(0) \\
\hline
   SMC \\
  \hline	   
NGC 104 (47 Tuc) &0 &0 &0.001(0) &0.002(0) &0.001(0) &0.001(0) &0 &0 &0 \\
NGC 362 &0.004(1) &0.009(1) & 0.01(0) & 0.01(0) &0.008(0) &0.005(0) &0.002(0) &0.001(0) &0 \\
\hline \hline
  \end{tabular}
  \caption{Probability of detecting at least one astrometric lensing event for each of the 9 IMBH masses considered, for $\tobs=20$, 25, and 30 years. 1-$\sigma$ error bars on the last decimal place are given in parentheses. An error bar of 0 indicates that the Poisson error is smaller than the precision quoted. \protect\label{tab:gcprob}}
  \end{center}
\end{table*}

In order to assess how to best exploit the available archival data, or how to devise the most efficient future long-term observing strategy to maximise chances of IMBH detections, we looked at how the probabilities given in \Tab{tab:gcprob} change for different values of $\tobs$. We considered values of 25 to 30 years and scaled the number of observations linearly with $\tobs$. The effect of $\tobs$ on the expected detection rate is shown in \Fig{fig:mass_ndet}. The expected number of detections rises approximately linearly, and at a rate that is faster for larger IMBH masses. For the lower mass ($< 500 \msun$) IMBHs, the increase is small: since $\te$ is smaller, even a relatively short observing campaign is sufficient to cover a significant portion of the astrometric curve, and the returns of extending the observing baseline are modest. On the other hand, for high-mass IMBHs, even a small increase in observing baseline improves event numbers significantly. For the four most promising clusters we have identified, adding even just five years to the baseline increases the probability of detecting an event significantly for the largest IMBH masses. For example, for M~22, compared to a 20-year campaign, for a $10^4 \msun$ IMBH, a baseline of $\tobs=25$ years increases $\avn$ by $\sim40\%$, raising the probability of detection from 0.44 to 0.56, while $\tobs=30$ years yields approximately double the expected detection rate, bringing the probability to 0.68.

We also looked at the impact of varying the number of observations between 1 and 5 epochs per year, and found that the improvement is not significant. In fact observing cadences down to 1 observation every $\sim 2-3$ years produce very similar probabilities, particularly for the slow events involving higher-mass IMBHs.

\section{Conclusions}\label{sec:conclusions}

There are many existing observations from various science programs that can already be used to search for the astrometric gravitational lensing signals caused by the presence of IMBHs in globular clusters. In addition to these, future observations that will be obtained for many clusters for a wide range of science objectives, in particular stellar population studies, will extend the time baseline of astrometric data sets for the clusters in our sample. The current available baseline for M~22 in the HST archive is 22 years, meaning that the expected number of detectable lensing events in the existing data set for a $10^4\msun$ IMBH is around $\avn=0.6$ for this cluster. This number rises to 1 if the baseline is extended by another 5 years, meaning that we have an excellent opportunity to make the first unambiguous detection of an IMBH, or to place stringent limits on the presence of IMBHs in the core of M~22.

Many globular clusters will continue to be observed by future facilities such as the \textit{James Webb Space Telescope} (JWST) and the \textit{Wide-field Infrared Survey Telescope} (WFIRST). These telescopes will be able to make astrometric measurements with precisions similar to or better than what can be achieved with HST, further extending our astrometric baseline throughout their mission lifetimes. This would allow for the detection of IMBHs in several globular clusters if they exist, and to obtain constraints on the demographics of these elusive objects.

\section*{Acknowledgements}

We thank Andrea Bellini for sharing with us his catalogues of proper motions for NGC 104, NGC 362, and NGC 6656, Manuela Zoccali for sending us her proper motion catalogue of NGC 6553, Luigi Bedin for sharing his proper motion and photometry catalogue of M~4, and Cristina Pallanca for her photometric catalogue of M~28. NK thanks Murray Brightman for useful discussions.

\bibliographystyle{mn2e}
\bibliography{../thesisbib}
\bsp

\label{lastpage}

\end{document}